\documentclass[journal]{IEEEtran}
\usepackage{cite}
\usepackage{amsmath,amssymb,amsfonts}
\usepackage{algorithmic}
\usepackage{graphicx}

\usepackage{dutchcal}
\DeclareMathAlphabet{\dutchcal}{OMS}{zplm}{m}{n}

\makeatletter
\def\endthebibliography{%
	\def\@noitemerr{\@latex@warning{Empty `thebibliography' environment}}%
	\endlist
}
\makeatother

\begin{document}

\title{Iterative Interference Cancellation\\ for Time Reversal Division Multiple Access}
\author{Ali Mokh, \IEEEmembership{Member,~IEEE}, George C. Alexandropoulos, \IEEEmembership{Senior~Member,~IEEE},\\ Mohamed Kamoun, \IEEEmembership{Member,~IEEE}, Abdelwaheb Ourir, Arnaud Tourin, Mathias Fink, and Julien de Rosny
\thanks{This work has received support under the program ``Investissements d’Avenir'' launched by the French Government and the Huawei Innovation Research Program.}
\thanks{A. Mokh, A. Ourir, A. Tourin, M. Fink and J. de Rosny are with the Institut Langevin, ESPCI Paris, Université PSL, CNRS, 75005 Paris, France (e-mails: \{ali.mokh, abdelwaheb.ourir, arnaud.tourin, mathias.fink, julien.derosny\}@espci.psl.eu)}
\thanks{G.~C.~Alexandropoulos is with the Department of Informatics and Telecommunications, National and Kapodistrian University of Athens, 15784 Athens, Greece (e-mail: alexandg@di.uoa.gr)}
\thanks{M. Kamoun is with the Mathematical and Algorithmic Sciences Lab, Paris Research Center, Huawei Technologies France, 92100 Boulogne-Billancourt, France (e-mail: mohamed.kamoun@huawei.com)}
}

\maketitle

\begin{abstract}
Time Reversal (TR) has been proposed as a competitive precoding strategy for low-complexity devices, relying on ultra-wideband waveforms. This transmit processing paradigm can address the need for low power and low complexity receivers, which is particularly important for the Internet of Things, since it shifts most of the communications signal processing complexity to the transmitter side. Due to its spatio-temporal focusing property, TR has also been used to design multiple access schemes for multi-user communications scenarios. However, in wideband time-division multiple access schemes, the signals received by users suffer from significant levels of inter-symbol interference as well as interference from uncoordinated users, which often require additional processing at the receiver side. This paper proposes an iterative TR scheme that aims to reduce the level of interference in wideband multi-user settings, while keeping the processing complexity only at the transmitter side. The performance of the proposed TR-based protocol is evaluated using analytical derivations. In addition, its superiority over the conventional Time Reversal Division Multiple Access (TRDMA) scheme is demonstrated through simulations as well as experimental measurements at $2.5$ GHz carrier frequency with variable bandwidth values.
\end{abstract}




\section{Introduction}
Ultra-WideBand (UWB) waveforms have ushered in a new era in short-range wireless sensor networks, enabling both robust communications and accurate ranging capabilities \cite{zhang2009uwb}. One central issue that UWB waveforms face when applied to low complexity devices is the effective energy collection, particularly when it is dispersed in rich multipath channels. TR, which belongs to spatio-temporal wavefront shaping techniques\cite{Lerosey2022} appears to be a paradigm shift in exploiting rich multipath conditions to enable the focus of UWB pulses in both time and space \cite{lerosey2004time}. Basically, TR consists of probing, flipping and transmitting back the channel responses. To probe them, one can use either classical channel estimation techniques or impedance modulation ones \cite{Yeo2022}. As for the flipping operation, it is performed by the transmitter either directly on the digitalized signals or by an analog device\cite{Wang2019a}.

Because the processing complexity is shifted to the transmitter side, TR allows receivers to benefit an high level of signal-to-noise ratio (SNR) via the use of simple, low cost, and low power coherent or even non-coherent reception \cite{wang2011green}. The temporal pulse compression offered by TR contributes to mitigating Inter-Symbol Interference (ISI) resulting from multipath signal propagation, while TR's spatial focusing capability can reduce Inter-User Interference (IUI) \cite{oestges2004time}. When the compression effect is sufficiently high, TR can be used as physical security layer\cite{Golstein2020}. 

From a communication theory perspective, TR is a precoding technique that can be used both for Single-Input Single-Output (SISO) and Multiple-Input Single-Output (MISO) impulse radio UWB wireless communication systems \cite{liu2008performance,Wang2019a,Wang2019}. Ten years ago, Time Reversal Division Multiple Access (TRDMA) was introduced to spatiotemporally focus messages to different users \cite{han2012time}, i.e., for multi-user MISO configurations. This technique provides a cost-effective waveform for a multiple access strategy, enabling energy efficient Internet of Things (IoT) nodes \cite{wang2011green}. Its implementation requires basic signal processing, enabling the network and, in particular, base stations to support numerous simultaneous downlink and uplink connections with a moderate complexity \cite{han2014multiuser}. In \cite{wang2016snr}, a theoretical analysis for the average SNR of the intended and unintended receivers in a distributed TR transmission scheme was presented, while the authors in \cite{lei2019performance} devised a probabilistic analysis of the SNR at the receiver in TR-based communication systems. Experimental evaluations of TR were recently presented in \cite{mokh2021indoor,mokh2021time,mokh2022freq273,TR_mag2022} for sub-6GHz and sub-THz communications.

Interference is one of the main factors that affect the performance of TR precoding. An iterative algorithm to reduce the ISI level in the received signal after TR precoding was presented in \cite{montaldo2004telecommunication} for Multiple-Input Multiple-Output (MIMO) communications. However, the presented algorithm is based on iterative feedback between the transmitter and the receiver, which makes the technique non-applicable for very low duty cycle nodes (including various IoT devices). In \cite{mokh2022time}, an alternative approach for Iterative TR (ITR) algorithms was designed according to which, only the first channel estimation is used to compute the focusing signal. The performance of the proposed technique was evaluated in a mobile scenario. In this paper, we introduce an important improvement of the ITR algorithm, where we annihilate  interference for a predetermined rate-back off. The motivation for this approach is, first, to avoid the handshake between the transmitter and the receivers, and second, to tune the complexity of the transmitter with the number of iterations allowed for each receiver. We evaluate the efficiency of the proposed ITR algorithm to cancel ISI and IUI in UWB channels both by computer simulations and via a novel experimental setup with a high speed arbitrary generator and a digital sampling oscilloscope at $2.5$ GHz carrier frequency with up to $100$ MHz communication bandwidth. In our performance evaluations, we have compared the proposed ITR technique with a reference TR-based algorithm for interference cancellation as well as the Regularized Zero-Forcing (RZF) technique \cite{heath2018foundations}, showcasing the superiority of our proposed algorithm for low complexity devices, such as IoT.

The rest of the paper is organized as follows. In the next Section, the principle of TRDMA is introduced and the proposed ITR technique is described. In Section III, we present the concept of equivalent channel for both conventional TR and ITR to evaluate the impact of ISI and IUI on the performance of the studied algorithms. The Symbol-to-Interference-plus-Noise Ratio (SINR) is also assessed for single-tap receivers in multipath Rayleigh channels. In Section IV, ITR is compared to the Regularized Zero-Forcing (RZF) algorithm in terms of SINR and computational complexity. An experimental multi-user MISO implementation of ITR is presented in Section V. The conclusions of the paper are included in the last Section VI.
%
\section{Time Reversal for Multiple Access}
\label{SEC:TRDMA}
%
\subsection{Time Reversal Division Multiple Access (TRDMA)}
We consider a multi-user MISO wireless communication system comprising one access point that is equipped with $M$ antenna elements and wishes to communicate in the downlink direction with $N$ single-antenna users. The baseband Channel Impulse Response (CIR) at a discrete time $k$ between the $m$-th ($m=1,2,\ldots,M$) transmitting antenna and the $i$-th ($i=1,2,\ldots,N$) receiving user/antenna is represented by $h_{i,m}[k]$. TR precoding utilizes the time reversed CIR, i.e., $h^*_{i,m}[L-k]$ with $L$ denoting the number of the significant channel taps, to focus the information-bearing electromagnetic field on the $i$-th receiving antenna. In mathematical terms, the TR-precoded signal sent by each $m$-th transmit antenna in order to focus each baseband information message $x_i[\mathcal{l}]$, with $\mathcal{l}=1,2,\ldots,\mathcal{P}$, on its respective $i$-th receiving antenna/user is given by the following expression:
\begin{equation}
     s_m[k]=\sum_{i=1}^N  \frac{\sum_{\mathcal{l}=1}^\mathcal{P} x_i[\mathcal{l}] h_{i,m}^*[L+\mathcal{l}D-k]}{\sqrt{\sum_{m'=1}^M\sum_{l=1}^{L} \lvert h_{i,m'}[l] \rvert^2}}, 
     \label{eq:yi}
\end{equation}
where $D$ denotes the rate back-off, i.e., the tap interval between two consequent symbols, which can be larger than one channel tap. Note that in practice, $h_{i,m}[k] \neq 0$ only when $k$ lies between  0 and  $L$.  The time interval $T_s$ between two consequent symbols is $D$ times the duration of a single tap $T_p$, i.e., the inverse of the system bandwidth $B$. It is noted that the normalization factor in \eqref{eq:yi} ensures that the power emitted toward each single-antenna user is the same. 

The signal received at each $j$-th ($j=1,2,\ldots,N$) is the sum over the $M$ emitting antennas of the convolution of $s_m[k]$ and the CIR $h_{j,m}[k]$. Assuming an Additive White Gaussian Noise (AWGN) channel, the received signal at each $\mathcal{k}$-th time interval can be mathematically expressed as follows:  
\begin{equation}
   y_j[\mathcal{k}]= \sum_{\mathcal{l}=1}^\mathcal{P} \sum_{i=1}^N x_i[\mathcal{l}] R_{j,i}[D(\mathcal{k}-\mathcal{l})-L]+n_j[\mathcal{k}],
   \label{eq:TR}
\end{equation}
where $R_{j,i}[k]$ represents the correlation function between the CIRs of the $j$-th and $i$-th receiving users/antennas, which is given as follows: 
\begin{equation}
 	R_{j,i}[k] = \frac{\sum_{k'=1}^L\sum_{m=1}^M h_{i,m}^*[k'] h_{j,m}[k'+k]}{\sqrt{\sum_{m'=1}^M\sum_{l=1}^L \lvert h_{i,m'}[l] \rvert^2}}.
   \label{eq1}
\end{equation}
In Eq.~(\ref{eq:TR}),  $n_j[\cdot]$ denotes the AWGN at the $j$-th receiver having standard deviation $\sigma$. Note that, in rich scattering wireless communication channels (i.e., Rayleigh fading conditions), when applying TR precoding with the CIR toward the $i$-th receiving antenna, i.e., $j=i$, the time-reversed field focuses in time and in space at this antenna, exhibiting a real-valued signal peak centered at tap $k=0$; in this case, all $M \times L$ signals add up coherently. The amplitude of this peak is given by the CIR's autocorrelation function at lag $0$, i.e., by $R_{i,i}[0]$. However, this focusing effect of TR is surrounded by side lobes due to the incoherent sum in~(\ref{eq1}), which produces on one hand inter symbol interferences (ISI) due to the fact $R_{i,i}[D\mathcal{k}] \neq 0$ for  $\mathcal{k} \neq 0$ and on the other hand inter user interferences (IUI) because $R_{j,i}[D\mathcal{k}] \neq 0$ when $ j\neq i$.

In the following, we propose an iterative algorithm that requires only one estimation of the MIMO channel matrix, irrespective of the number of iterations.

\subsection{Proposed Iterative Algorithm}
 Iterative deconvolution processing is popular because of its robustness~\cite{ligorria1999iterative}. Here, we propose to adapt such an algorithm to iteratively compute the signal $g_{i,m}[k]$ to be emitted by the $m$-th transmit antenna to focus a peak on user $i$ at tap $k=L$. This removes the secondary peaks only at the symbol times, i.e., at the channel taps $\mathcal{k} D$ on user $i$ and on the other users. The flow chart of this iterative time reversal (ITR) algorithm is illustrated in~Fig.~\ref{fig:algo}. In the chart, we have used the notation $\bar{h}_{i,m}[k]$ for the normalized CIR, i.e., $\bar{h}_{i,m}[k] \triangleq h_{i,m}[k]/\sqrt{\sum_{k'=1}^L\sum_{m'=1}^M \lvert h_{i,m'}[k'] \rvert^2}$. The algorithmic steps of ITR are summarized as follows:
\begin{figure}[t!]
\centering
\includegraphics[width=0.9\linewidth]{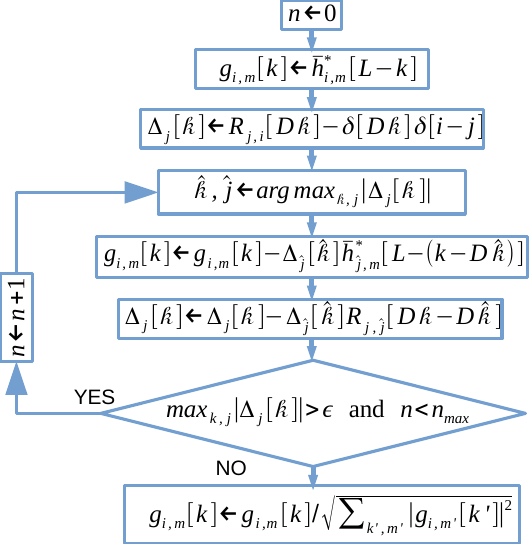}
\caption{Flow chart of the proposed ITR algorithm for multi-user MISO communication systems.  Index $n$ indicates the number of the algorithm iteration and $n_{\rm max}$ the total allowable number of iterations, whereas $\epsilon$ is a constant used for evaluating the algorithm's performance, and thus, determines its termination. The algorithmic steps for implementing the TR precoded signal $g_{i,m}$ at each $m$-th transmit antenna to cancel out the spurious taps at every $D$ channel taps when received at $i$-th user are illustrated. The notation $a \leftarrow b$ stands for the assigment $a$ to $b$.}
\label{fig:algo} 
\end{figure}
\begin{figure}[t!]
\centering
\includegraphics[width=1 \linewidth]{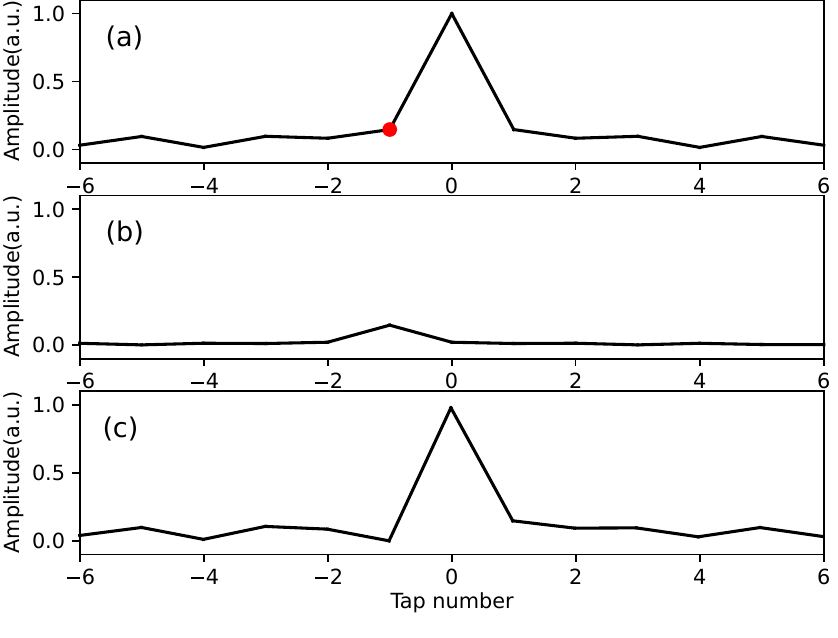}
\caption{Schematic representation of an iteration of the ITR algorithm, considering, for simplicity, only one receiver/user. (a) Modulus of the TR signal to remove spurious peaks. The red circle represents the maximum spurious contribution to cancel (this is accomplished via ITR's Step 2). (b) Another TR signal adequately shifted in time in order for its maximum peak to appear in the place of the maximum spurious contribution in (a) having the same amplitude with that (ITR's Step 3). (c) The modulus of the ITR's final signal that results from the difference between the signals in (a) and (b) (Step 4 of ITR). As observed, the maximum spurious contribution in (a) at the position of the circle has been removed.}
\label{fig:expli} 
\end{figure}
\begin{enumerate}
    \item The signal $g_{i,m}[k]$ is initiated with the CIR flipped in time as follows:
    \begin{equation}
        g_{i,m}[k] = {\bar h}_{i,m}^*[L-k].
    \end{equation}
	\item Using (\ref{eq:TR}) with $n_j[\mathcal{k}]=0$, the focused signal at the symbol taps is computed together with the difference signal $\Delta_i[\mathcal k]$ between the aforementioned signal and the ideal focusing (i.e., the signal $\delta[D \mathcal k]\delta[i-j]$); this is illustrated in~Fig.~\ref{fig:expli}(a).
	\item The received antenna/user $\hat j$ and the tap $\hat{\mathcal{k}} D$ where the maximum difference occurs are found. This corresponds to the highest (undesired) secondary peak at the taps being multiples of $D$ (see~Fig.~\ref{fig:expli}(a)).
	\item An inverted TR signal $-{\bar h}_{\hat j,m}^*[L-k + \hat k]$ is generated to focus a pulse on this secondary tap on the $\hat j$-user. This pulse has an opposite sign and the proper scale amplitude $\Delta_{\hat j}[\hat {\mathcal k}]$ (see~(Fig.~\ref{fig:expli}(b)). 
	\item The aforementioned secondary peak is canceled using the inverted TR pulse, as shown in~Fig.~\ref{fig:expli}(c). As shown in this figure, the time-shifted and scaled CIR to obtain the inverted pulse is added to the signal to be transmitted to cancel the highest secondary tap.
	\item The difference signal $\Delta_{j}[\mathcal k]$ is finally updated.
\end{enumerate}
The repetition of the ITR algorithm from Step 2 to Step 6 removes progressively the secondary taps in the focused signal. The algorithmic iteration stop either when a given number $n_{\rm max}$ of maximum iterations is reached or when $\left|\Delta_{j}[\mathcal k]\right|$ becomes smaller than a target value $\epsilon$. The former option can be used to tune ITR's complexity and the processing latency at the transmitter side, while the latter guarantees that the algorithms terminates when a required SNR level is met. It noted that, as shown in the last algorithmic step in Fig.~\ref{fig:algo}, the ITR precoded signal is normalized in energy before transmission.

\subsection{Equivalent Channel and SINR Definition}
Let ${f}^{[D]}$ represent the equivalent channel after TR precoding at a rate back-off $D$. By assuming a computationally simple receiver performing only signal synchronization, that is based on the signal peaks, and being capable to process only one sample per symbol, the post-processed received signal can be written as follows:
\begin{equation}
y_j[\mathcal{l}]= \sum_{\mathcal{k}=1}^{\mathcal{P}}\sum_{i=1}^N f^{D}_{j,i}[\mathcal{l}-\mathcal{k}]  x_i[\mathcal{k}] + n_j[\mathcal{l}],
\end{equation}
where we have used the following notation for the case of conventional TR precoding:
\begin{equation}
{f}^{D}_{j,i}[\mathcal{l}] = {R_{j,i}[-D\mathcal{l}-L]}.
\end{equation}
In the case of proposed ITR algorithm, the equivalent channel ${f}^{D}_{j,i}[l]$ can be expressed as follows:
\begin{equation}
{f}^{D}_{j,i}[\mathcal{l}] =\sum_{k} \sum_{m=1}^M g_{i,m}[D\mathcal{l}-k] h_{j,m}[k].
\end{equation}
From these equivalent channel approach (i.e., for conventional TR or ITR), the ISI and IUI at each $i$-th user can be computed as:
    \begin{equation}
        \textrm{ISI}_{i} = \mathbb{E}\{x^2\} \sum_{\mathcal{k}=-L/D ,\,\mathcal{k} \neq 0}^{L/D} \left| f^{D}_{i,i}[\mathcal{k}] \right|^2,
    \end{equation}
    and 
      \begin{equation}
        \textrm{IUI}_{i} = \mathbb{E}\{x^2\} \sum_{j=1,\,j \neq i}^N \sum_{\mathcal{k}=-L/D}^{L/D} \left| f^{D}_{j,i}[\mathcal{k}] \right|^2,
    \end{equation}
    where $\mathbb{E}\{x^2\}$ is the mean power level of the transmitted symbols. As a consequence, the expression of SINR is given by:
\begin{equation}
\mbox{SINR}_i = \frac{|f^{D}_{i,i}[0]|^2}{\sum_{\mathcal{k}\neq 0} |f^{D}_{i,i}[\mathcal{k}]|^2+ \sum_{j\neq i}\sum_{\mathcal{k}=1} |f^{D}_{j,i}[\mathcal{k}]|^2  + \sigma^2}.
\label{eq:SINR}
\end{equation}

In the following, the SINR metric will be used to assess the performance of the proposed ITR algorithm.

\section{Simulation Results}
In this section, we evaluate the performance of the proposed ITR algorithm, focusing on the investigation of its interference cancellation capability. 

\subsection{Channel Model}
The performance has been assessed via simulations on synthetic channels following the Rayleigh distribution with exponential power decay time. In particular, the channels were drawn from multi-tap realizations with amplitudes decaying exponentially with time $\tau$. The probability density function of the CIR between each $i$-th transmit antenna and $m$-th user was given by:
\begin{equation}
 \mathcal{P}(h_{i,m}[k]) = \mathcal{C} \mathcal{N}(0,1) \exp\left(-\frac{k}{2\tau}\right),
\end{equation}
with $\mathcal{C} \mathcal{N}(0,1)$ denoting the zero-mean and unit-variance complex normal distribution. For simplicity, we have assumed that the channel taps are decorrelated from one to another (i.e., independent channel realizations). 

\subsection{Interference Cancellation Performance}
\begin{figure}[t!]
\centering
\includegraphics[width=1 \linewidth]{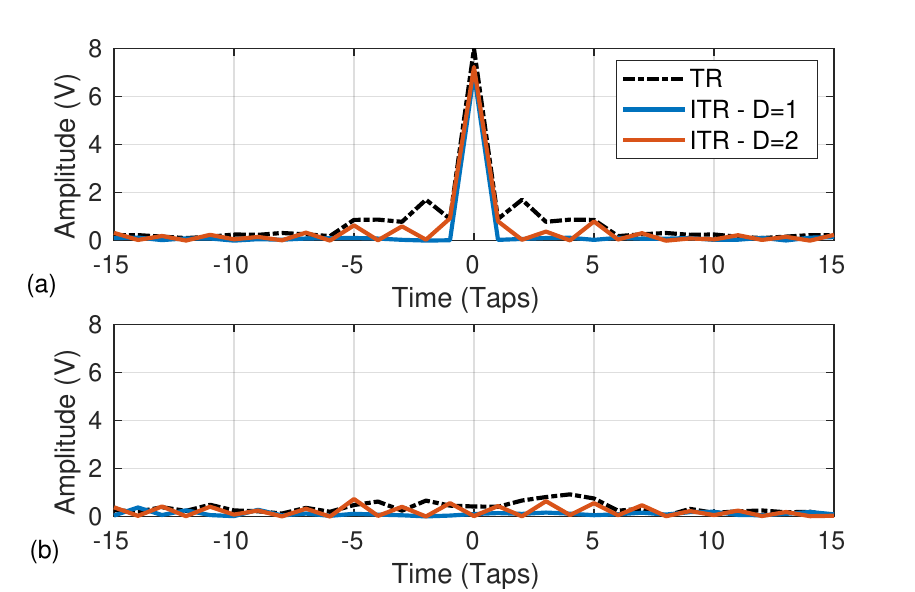}
\caption{Received signal strength in Volts versus time in terms of channel taps for conventional TR (black dashed line) and the proposed ITR focusing (blue and red solid lines), considering Rayleigh fading channels, a power decay time of $5$ taps, and two disctinct values for the rate back-off $D$ (blue line for $D=1$ and red line for $D=2$). The maximum number of iterations for ITR was set to $100$. (a) The received signal on user $1$ when focusing on it; and (b) The received signal on user $2$ when focusing on user $1$.}
\label{fig:focus} 
\end{figure}

Using the aforedescribed channel model, we next consider a simple scenario comprising one access point equipped with an $8$-element antenna array and two single-antenna users. Figure~\ref{fig:focus} compares the signal level at each receiving user between conventional TR precoding and the proposed ITR algorithm when using the rate back-off values $D=1$ and $2$. It can be seen in the figure that the ITR approach reduces drastically the level of spurious contributions, while reducing slightly (limited to $10\%$) the level of the desired signal peak. It also shown that the interference in all taps (except the central tap which needs to be kept) is reduced for $D=1$, while for $D=2$, only the interference at each $2$ successive taps is removed. 

\begin{figure}[t!]
\centering
\includegraphics[width=1 \linewidth]{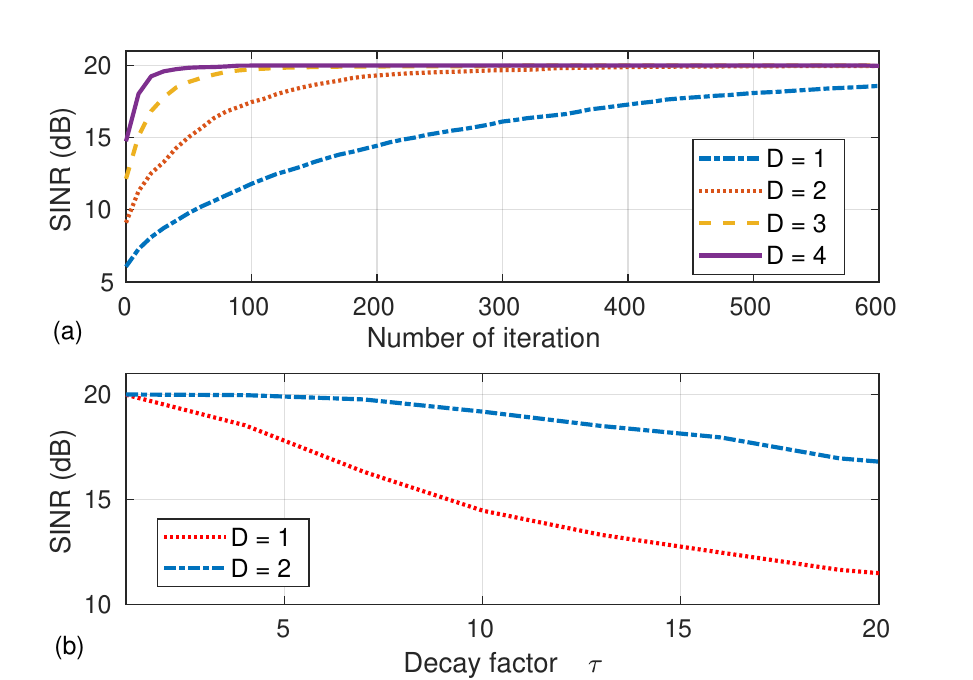}
\caption{SINR performance in dB for the ITR algorithm: (a) with respect to the number of algorithmic iterations, considering a Rayleigh fading channel with spreading time $\tau=5$ taps and $D=\{1,2,3,4\}$; and (b) with respect to the spreading time of the Rayleigh channel, considering $200$ iterations and $D=\{1,2\}$.}
\label{fig:SINR} 
\end{figure}

To quantify the improvement brought by the proposed ITR algorithm, we have set the noise at the receivers to a fixed level, so that the ratio between the transmit signal power and the Gaussian noise at each receiver is $20$ dB, i.e., we have set $\sigma =0.1$ in~(\ref{eq:SINR}). 
In Fig.~\ref{fig:SINR}(a), the SINR performance in dB is sketched as a function of the number of ITR algorithmic iterations, considering a power decay time of $5$ taps and different values for $D$. A strong increase in the SINR is observed with increasing numbers of iterations, and interestingly, convergence is achieved to the maximum value $20$ dB, which is the link's SNR level, as previously described. As demonstrated, for higher $D$ values, a higher SINR is obtained for the same number of iterations. It is also shown that the SINR performance reaches its maximum value after $50$ iterations when $D=4$, while $120$ iterations are required for $D=3$, $350$ iterations for $D=2$, and more than $600$ iterations for $D=1$. The dependence of the SINR with the decay factor $\tau$ after $200$ iterations of the ITR algorithm for $D=1$ and $2$ is depicted in Fig.~\ref{fig:SINR}(b). It can be seen that the SINR decreases with $\tau$. In fact, long CIRs introduce more ISI and IUI, which requires more algorithmic iterations to be removed. As observed, the slope of the decrease in the SINR curve with increasing $\tau$ values is faster for $D=1$ than for $D=2$. This is due to the fact that more interference taps need to be removed when $D$ gets smaller values.

\section{Comparison of ITR and RZF Precoding} 
For the comparison between the multi-user precoding techniques ITR and RZF, we have considered the following two metrics: \textit{i}) the signal and interference power levels; and \textit{ii}) the computational complexity. The RZF was implemented in the frequency domain at each subcarrier, hence, Fast Fourier Transform (FFT) was first applied to transform the time-domain signal in the frequency domain. The RZF expression for the $M\times N$ complex-valued precoding matrix $\textbf{W}(f)$ at each utilized subcarrier $f$ is given by:
\begin{equation}\label{eq:RZF_fd}
    \textbf{W}(f)\triangleq \textbf{H}^H(f)\left(\textbf{H}^H(f)\textbf{H}(f)+\alpha \textbf{I}\right)^{-1},
\end{equation}
where $\textbf{H}(f)$ represents the $N \times M$ channel gain matrix at subcarrier $f$, $\alpha$ is the regularization factor ($\alpha=0$ gives the standard ZF), and $\mathbf{I}$ denotes the $M\times M$ identity matrix. The RZF precoder in the time domain is finally obtained from \eqref{eq:RZF_fd} by applying the inverse FFT. For the comparisons that follow, we have have used a synthetic Rayleigh fading channel and a large number of independent channel realizations. For the fairness of the comparison, we insured that the average transmit power is the same when using RZF and ITR, and we calculated the power of the received signal as well as that of interference when using each of the three multi-user precoding schemes. 

\subsection{Signal and Interference Power Levels}
Table~\ref{SIR_measures} summarizes the received signal (S) and interference (I) power levels in dB for the two considered multi-user precoders and their extreme versions, i.e., ITR and its conventional TR version as well as RZF and its simple ZF version. An access point with an antenna array of $M=2$ and $4$ elements was considered. As shown in the table, the performance of all precoders improves when the number of transmit antenna increases. It is also depicted that the degradation of the received signal level with ITR, as compared to TR, is less than $2$ dB for both $10$ and $20$ algorithmic iterations. The same order of level drop is observed with ZF precoding. The latter witnesses that, although ZF precoding completely nulls interference, this comes with the cost of strong attenuation in the desired received signal. Interestingly, ITR results in a stronger desired signal. The table evidently indicates that RZF with $\alpha=0.3$ outperforms both ZF and ITR in terms of the compromise between interference reduction and attenuation of the desired signal. However, ITR is based on simple signal processing operations, hence, it is easier to implement requiring less computational resources than RZF. To quantify the latter statement, in the sequel, we present the computational complexities of ITR and RZF.
\subsection{Computational Complexity}
The first task to compute the $g_{i,m}[k]$ (see~Fig.~\ref{fig:algo}) is to evaluate the correlation function  $R_{j,i}[k]$. It requires the convolution of $M N^2$ of pairs of CIRs (each direct linear convolution requires $L^2$ scalar multiplications). Alternatively, the linear convolutions in time can be performed in the frequency domain as scalar multiplications; this needs $M N^2 L$ scalar multiplications and another $2 L \log_2(L)$ scalar multiplications to perform the FFT and inverse FFT operations. The task of the iterative steps to perform interference cancellation requires $M L$ scalar multiplications per iteration, resulting in a total of $n' M L$ operations (without loss of generality, we here assume that $n'$ denotes the total number of ITR's iterations for any of its termination options).

On the other hand, to perform the RZF precoding, the following complexity is needed:
\begin{itemize}
    \item $2L \log_2(L)$ multiplications for the FFT and inverse IFFT operations; and
    \item $N^3 + MN$ operations for the matrix inversion at each subcarrier $f$, resulting in a total of $L(N^3 + MN)$ operations for an $L$-tap channel.
\end{itemize}

The total computational complexity of the ITR and RZF schemes is presented in Table~\ref{complexity}. It is clear that ITR is more computationally efficient than RZF, with this superiority increasing as the number of users $N$ increases.
\begin {table}[t]
\caption{Signal (S) and interference (I) power levels in dB for different versions of the four considered multi-user precoding schemes (ZF, RZF, TR, and ITR) for the case of an access point with $M=\{2,4\}$ antennas.}
\label{SIR_measures}
\begin{center}
\begin{tabular}{|c|c|c|c|c|}
\hline
Number of Antennas $M$ &  \multicolumn{2}{c}{2} &  \multicolumn{2}{|c|}{4} \\
\hline
Precoding Scheme & S [dB] & I [dB] & S [dB] & I [dB] \\
\hline\hline
ZF & -7.36 & $-\infty$ & 3.25 & $-\infty$\\
\hline
RZF ($\alpha=$0.1) & -0.26 & $-10.4$ & 4.24 & -13.9\\
\hline
RZF ($\alpha=$0.3) & 1.23 & $-5.63$ & 4.94 & -7.54\\
\hline
TR & 3 & 2.65 & 6.02 & 2.68\\
\hline
ITR (10 iterations) & 1.6 & 0 & 5.1 & -0.61\\
\hline
ITR (20 iterations) & 1.17 & -1.24 & 4.89 & -2.4\\
\hline
\end{tabular}
\end{center}
\end {table}
\begin {table}[t!]
\caption{Computational complexity comparison between ITR and RZF.}
\label{complexity}
\begin{center}
\begin{tabular}{|c|c|}
\hline

ITR &  $n' ML+M N^2 L^2$ or $n'ML+M N^2 L (1 + 2log_2(L))$ \\
\hline
RZF& $2L\log_2(L)+L(N^3+MN)$\\
\hline
\end{tabular}
\end{center}
\end {table}

%
\section{Experimental Results}
In this section, we present novel experimental results for the proposed ITR technique, considering a multi-user MISO configuration in an indoor laboratory environment.

\subsection{Experimental Setup}
\begin{figure}[t!]
\centering
\includegraphics[width=1 \linewidth]{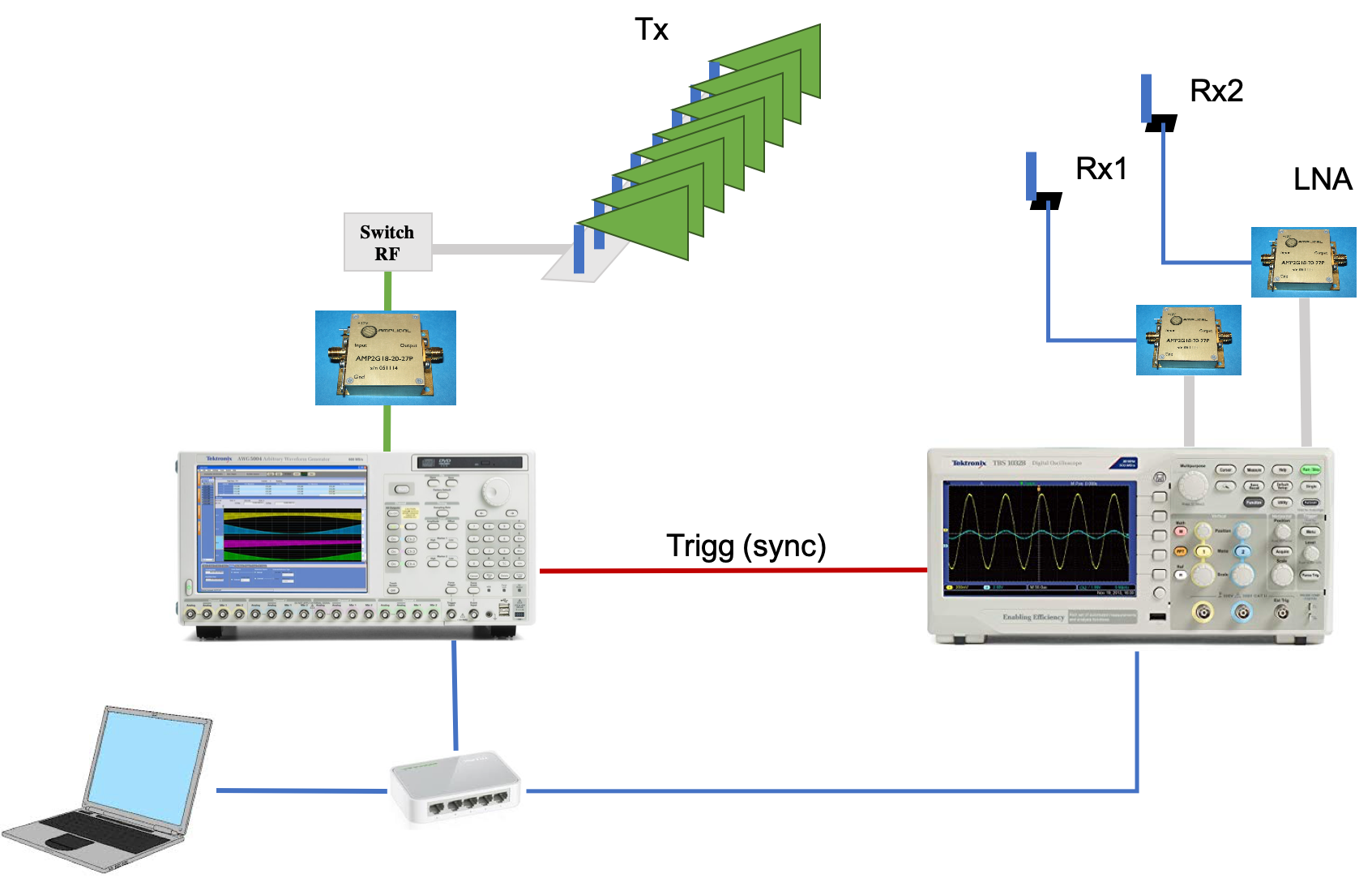}
\caption{The designed experimental setup, comprising one multi-antenna transmitter (Tx) and two single-antenna receivers (Rx1 and Rx2), for assessing the performance of the proposed ITR precoding scheme. LNA stands for Low Noise Amplifier and RF for Radio frequency, while the Trigg (sync) red line indicates the synchronization mechanism between the communication ends.}
\label{fig:Setup}
\end{figure}
The schematic of the devised experimental setup for realizing TR and ITR wireless communications is shown in Fig$.$~\ref{fig:Setup}. 
The transmitted signals were generated by a Tektronix AWG-7012 (for up to $10$ GSps signals) Arbitrary Waveform Generator (AWG), and then passed through a  Radio Frequency (RF) power amplifier before their over-the-air transmission. At the receiver side, the reception antennas were connected to a Low Noise Amplifier (LNA), whose output fed a Tektronix TDS6604B RF oscilloscope (for up to $12.5$ GSps). The AWG and oscilloscope were synchronized through a $100$ MHz reference signal and controlled via Ethernet by a desktop computer. 
We have considered a linear array at the transmitter (Tx) composed of $N_t=8$ Vivaldi antennas, all connected to a solid-state RF switch, whose input was fed with the RF power amplifier. Two dipole antennas were used at the receiver side, each connected to its own LNA and its own channel of the RF oscilloscope. The two receive antenna emulated the users Rx1 and Rx2. The experiment took place inside a reverberating chamber of size 1.5~m $\times$ 1~m $\times$ 0.5~m to create rich multipath channels, which is essential for taking advantage of the TR approach.

The estimation of the CIR and TR precoding were performed as follows. First, a chirp signal of one-second duration spanning the frequency range $[f_c-B/2, f_c+B/2]$ was transmitted. To estimate the CIR, the desktop computer performed the deconvolution of the received signal with the chirp signal by correlating the received signal with the emitted chirp. Then, the CIRs were flipped in time and then resampled to fit the sampling frequency of the AWG. For the ITR technique, the estimated CIR was first downconverted to baseband to apply the proposed iterative algorithm in~Fig.~\ref{fig:algo}. The precoding signal was then up-converted and emitted. The carrier frequency of all conducted experiments was set to $f_{c}=2.5$ GHz and we considered various values for the bandwidth $B$.

\subsection{Performance Measurements}
\begin{figure}[t!]
\centering
\includegraphics[width=1 \linewidth]{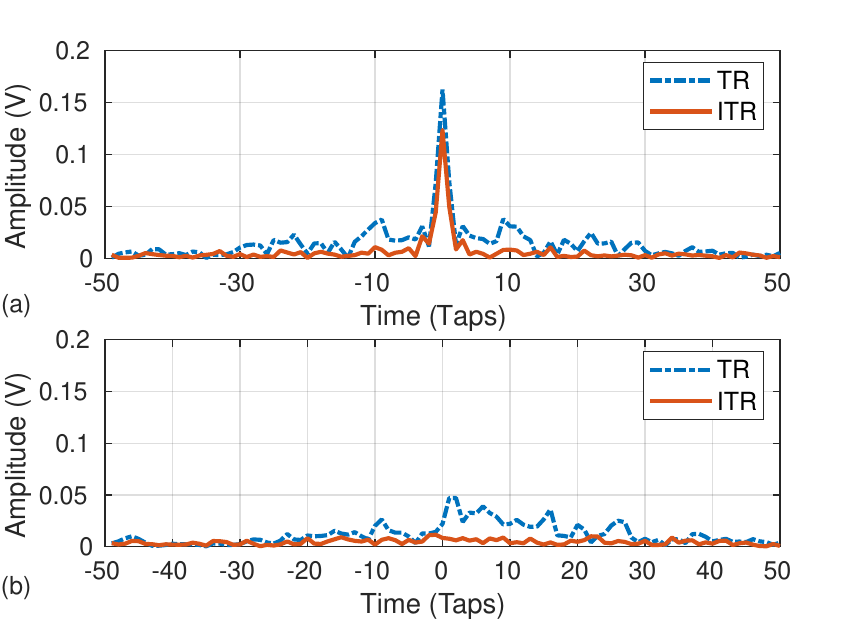}
\caption{Received signal strength measurements in Volts versus time in terms of channel taps for conventional TR (blue dashed line) and the proposed ITR focusing (red solid line). The maximum number of iterations for ITR was set to $400$. (a) The received signal on user $1$ when focusing on it; and (b) The received signal on user $2$ when focusing on user $1$. The bandwidth was set to $B=100$ MHz.}
\label{fig:TRVsITR}
\end{figure}

In the performance curves included in Fig.~\ref{fig:TRVsITR}, we have set $B=100$ MHz and implemented both conventional TR and the proposed ITR technique with a maximum of $400$ algorithmic iterations to focus the transmitted signal on Rx1. In Fig.~\ref{fig:TRVsITR}a, the baseband received signal at the targeted Rx1 is depicted, while Fig.~\ref{fig:TRVsITR}b includes the respective signal at Rx2. It can be observed that the ITR precoded signal results in slightly lower amplitude for the intended received signal at Rx1 at the time instant $t=0$, but also yield lower ISI and IUI, as compared to TR precoding. In addition, it can be seen from Fig.~\ref{fig:TRVsITR}b that with ITR precoding, the unintended signal at Rx2 is of lower amplitude than conventional TR. Thus, both subfigures indicate ITR's superiority in providing spatiotemporal focusing in the considered experimental multi-user scenario.

\begin{figure}[t!]
\centering
\includegraphics[width=1 \linewidth]{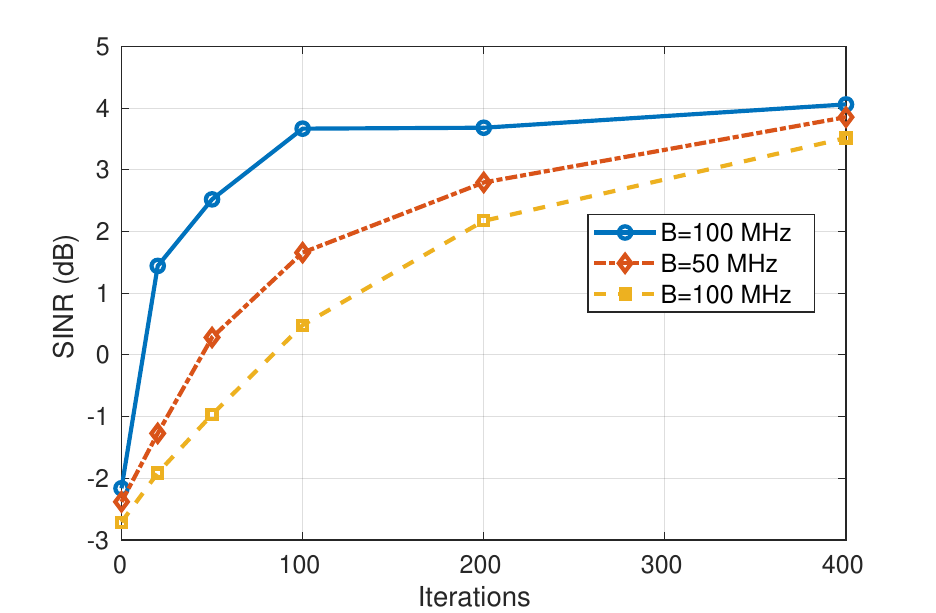}
\caption{SINR performance in dB for the ITR algorithm as a function of the number of its algorithmic iterations for $B=\{20,50,100\}$ MHz.}
\label{fig:SINRtoB}
\end{figure}

In Fig.~\ref{fig:SINRtoB}, considering the bandwidth values $B=20$, $50$, and $100$ MHz, we illustrate the SINR performance in dB, via expression~\eqref{eq:SINR}, with the proposed ITR technique versus the number of its algorithmic iterations. The SINR curves in the figure resulted from averaging over $12$ different configurations of the cavity, which were obtained with the help of a mechanical stirrer. It is shown that the SINR of the received signal increases with the number of iterations for all experimented bandwidth values. As observed, the lower the bandwidth, the faster (i.e., with fewer iterations) the SINR increases, and in particular: after $100$ iterations, the SINR increases from $-2$ to $3.8$ dB for $20$ MHz bandwidth; from $-2.5$ to $1.8$ dB for $50$ MHz bandwidth; and from $-2.8$ to $0.5$ dB for $100$ MHz bandwidth. This behavior is attributed to the fact that, for the same power decay time, the larger is the bandwidth, the more interference taps have to be removed. 

\begin{figure}[t!]
\centering
\includegraphics[width=1 \linewidth]{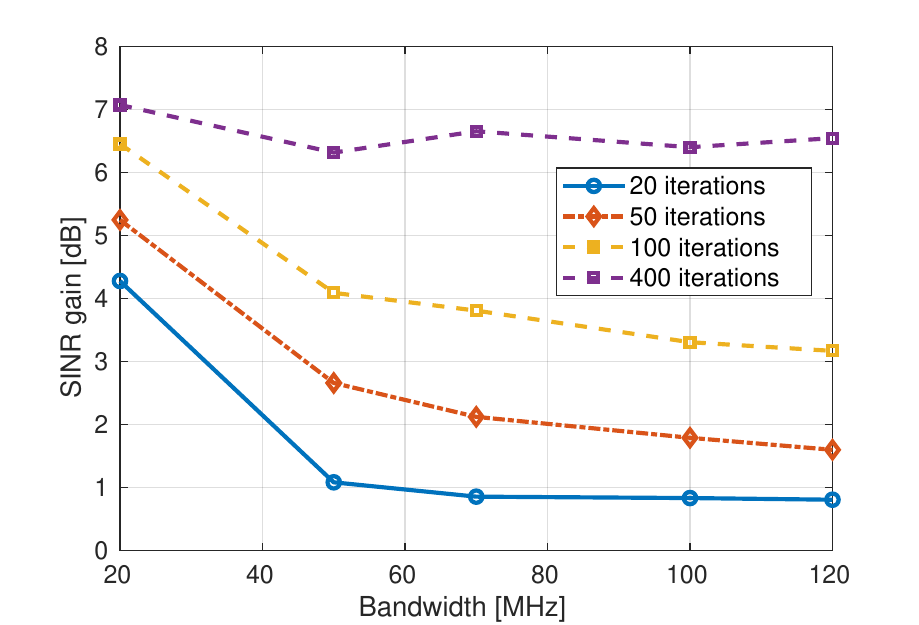}
\caption{SINR gain in dB of ITR, with respect to conventional TR, as a function of the bandwidth $B$ in MHz for different numbers of algorithmic iterations for ITR.}
\label{fig:GSINRtoB}
\end{figure}

Finally, in Fig.~\ref{fig:GSINRtoB}, the SINR gain in dB of ITR, with respect to conventional TR, is plotted versus the bandwidth values $B$ in MHz for different numbers of iterations for the proposed algorithm. It is depicted that, for $20$, $50$, and $100$ iterations, the gain decreases with increasing $B$. However, for $400$ iterations, the gain is approximately the same, and actually the largest, for all considered bandwidth values. For this case, the small fluctuations in the gain resulted from the sampling noise, which can be reduced if we run the experiment for much larger numbers of trials than we actually performed (i.e., $\gg12$ cavity configurations). The saturation of a SINR gain of 6.5dB-7dB is due to the imperfections of our setup (e.g., noise during the CIR estimation and the limited dynamic range of the digital sampler). 
%

\section{Conclusions}
In this paper, UWB-based TRDMA has been investigated. A novel ITR technique, that is based on an iterative TR-based mechanism, was presented that drastically reduces the effect of ISI and IUI in multi-user MISO communication systems. The proposed algorithm was extensively compared, in terms of signal and interference power levels as well as of the resulting computational complexity, with the RZF precoding. On one hand, it was demonstrated that ITR outperforms the basic ZF (i.e., without regularization), when small numbers of transmit antennas are used. On the other hand, when the regularization factor is introduced, RZF provides slightly better performance than ITR. However, this enhancement is achieved at the cost of a higher computational complexity. This led to the conclusion that ITR is suitable for low complexity devices (e.g., IoT devices) as compared to RZF. The performance of the proposed ITR technique was extensively evaluated in synthetic Rayleigh channels as well as through novel experimental investigations, considering different bandwidth values and numbers for the algorithmic iterations. It was showcased that ITR can increase the SINR performance in multi-user scenarios up to $7$ dB, as compared to conventional TR precoding. 

\bibliographystyle{IEEEtran}
\bibliography{ref}

\begin{thebibliography}{10}
\providecommand{\url}[1]{#1}
\csname url@samestyle\endcsname
\providecommand{\newblock}{\relax}
\providecommand{\bibinfo}[2]{#2}
\providecommand{\BIBentrySTDinterwordspacing}{\spaceskip=0pt\relax}
\providecommand{\BIBentryALTinterwordstretchfactor}{4}
\providecommand{\BIBentryALTinterwordspacing}{\spaceskip=\fontdimen2\font plus
\BIBentryALTinterwordstretchfactor\fontdimen3\font minus
  \fontdimen4\font\relax}
\providecommand{\BIBforeignlanguage}[2]{{%
\expandafter\ifx\csname l@#1\endcsname\relax
\typeout{** WARNING: IEEEtran.bst: No hyphenation pattern has been}%
\typeout{** loaded for the language `#1'. Using the pattern for}%
\typeout{** the default language instead.}%
\else
\language=\csname l@#1\endcsname
\fi
#2}}
\providecommand{\BIBdecl}{\relax}
\BIBdecl

\bibitem{zhang2009uwb}
J.~Zhang, P.~V. Orlik, Z.~Sahinoglu, A.~F. Molisch, and P.~Kinney, ``{UWB}
  systems for wireless sensor networks,'' \emph{Proc. IEEE}, vol.~97, no.~2,
  pp. 313--331, Feb. 2009.

\bibitem{Lerosey2022}
G.~Lerosey and M.~Fink, ``Wavefront shaping for wireless communications in
  complex media: From time reversal to reconfigurable intelligent surfaces,''
  \emph{Proceedings of the {IEEE}}, vol. 110, no.~9, pp. 1210--1226, sep 2022.

\bibitem{lerosey2004time}
G.~Lerosey, J.~De~Rosny, A.~Tourin, A.~Derode, G.~Montaldo, and M.~Fink, ``Time
  reversal of electromagnetic waves,'' \emph{Physical Review Lett.}, vol.~92,
  no.~19, p. 193904, 2004.

\bibitem{Yeo2022}
K.~B. Yeo, C.~Leconte, P.~del Hougne, P.~Besnier, and M.~Davy, ``Time reversal
  communications with channel state information estimated from impedance
  modulation at the receiver,'' \emph{{IEEE} Access}, vol.~10, pp.
  91\,119--91\,126, 2022.

\bibitem{Wang2019a}
Z.~Wang, B.-Z. Wang, D.~Zhao, and R.~Wang, ``Full analog broadband
  time-reversal module for ultra-wideband communication system,'' \emph{{IEEE}
  Photonics Journal}, vol.~11, no.~5, pp. 1--10, oct 2019.

\bibitem{wang2011green}
B.~Wang, Y.~Wu, F.~Han, Y.-H. Yang, and K.~R. Liu, ``Green wireless
  communications: A time-reversal paradigm,'' \emph{IEEE J. Sel. Areas
  Commun.}, vol.~29, no.~8, pp. 1698--1710, Sep. 2011.

\bibitem{oestges2004time}
C.~Oestges, J.~Hansen, S.~M. Emami, A.~D. Kim, G.~Papanicolaou, and A.~J.
  Paulraj, ``Time reversal techniques for broadband wireless communication
  systems,'' in \emph{Proc. European Microwave Conf.}, Amsterdam, Netherlands,
  Oct. 2004, pp. 49--66.

\bibitem{Golstein2020}
S.~Golstein, T.-H. Nguyen, F.~Horlin, P.~D. Doncker, and J.~Sarrazin,
  ``Physical layer security in frequency-domain time-reversal {SISO} {OFDM}
  communication,'' in \emph{2020 International Conference on Computing,
  Networking and Communications ({ICNC})}.\hskip 1em plus 0.5em minus
  0.4em\relax {IEEE}, feb 2020.

\bibitem{liu2008performance}
X.-F. Liu, B.-Z. Wang, S.-Q. Xiao, and J.~H. Deng, ``Performance of impulse
  radio {UWB} communications based on time reversal technique,'' \emph{Progress
  Electromagn. Research}, vol.~79, pp. 401--413, 2008.

\bibitem{Wang2019}
Z.~Wang, B.-Z. Wang, Y.~Wu, and D.~Zhao, ``A novel ultra-wideband communication
  system using an analog time-reversal module,'' in \emph{2019 {IEEE}
  International Symposium on Antennas and Propagation and {USNC}-{URSI} Radio
  Science Meeting}.\hskip 1em plus 0.5em minus 0.4em\relax {IEEE}, jul 2019.

\bibitem{han2012time}
F.~Han, Y.-H. Yang, B.~Wang, Y.~Wu, and K.~R. Liu, ``Time-reversal division
  multiple access over multi-path channels,'' \emph{IEEE Trans. Commun.},
  vol.~60, no.~7, pp. 1953--1965, Jul. 2012.

\bibitem{han2014multiuser}
F.~Han and K.~R. Liu, ``A multiuser {TRDMA} uplink system with {2D} parallel
  interference cancellation,'' \emph{IEEE Trans. Commun.}, vol.~62, no.~3, pp.
  1011--1022, Mar. 2014.

\bibitem{wang2016snr}
L.~Wang, R.~Li, C.~Cao, and G.~L. St{\"u}ber, ``{SNR} analysis of time reversal
  signaling on target and unintended receivers in distributed transmission,''
  \emph{IEEE Trans. Commun.}, vol.~64, no.~5, pp. 2176--2191, May 2016.

\bibitem{lei2019performance}
W.~Lei and L.~Yao, ``Performance analysis of time reversal communication
  systems,'' \emph{IEEE Commun. Lett.}, vol.~23, no.~4, pp. 680--683, Apr.
  2019.

\bibitem{mokh2021indoor}
A.~Mokh, R.~Khayatzadeh, J.~de~Rosny, M.~Kamoun, A.~Ourir, and A.~T.~M. Fink,
  ``Indoor experimental evaluation of ultra-wideband {MU}-{MISO} {TRDMA},'' in
  \emph{Proc. IEEE Vehicular Technology Conference (VTC2021-Spring)}, May 2021,
  pp. 1--5.

\bibitem{mokh2021time}
A.~Mokh, J.~De~Rosny, G.~C. Alexandropoulos, R.~Khayatzadeh, A.~Ourir,
  M.~Kamoun, A.~Tourin, and M.~Fink, ``Time reversal precoding at {SubTHz}
  frequencies: {E}xperimental results on spatiotemporal focusing,'' in
  \emph{Proc. IEEE Conference on Standards for Communications and Networking
  (CSCN)}, Dec. 2021, pp. 78--82.

\bibitem{mokh2022freq273}
A.~Mokh, J.~De~Rosny, G.~C. Alexandropoulos, M.~Kamoun, A.~Ourir,
  R.~Khayatzadeh, A.~Tourin, and M.~Fink, ``Experimental validation of time
  reversal multiple access for {UWB} wireless communications centered at the
  $273.6$ {GHz} frequency,'' in \emph{Proc. IEEE Vehicular Technology
  Conference (VTC2022-Spring)}, Helsinki, Finland, Jun. 2022, pp. 1731--1736.

\bibitem{TR_mag2022}
G.~C. Alexandropoulos, A.~Mokh, R.~Khayatzadeh, J.~De~Rosny, M.~Kamoun,
  A.~Ourir, A.~Tourin, M.~Fink, , and M.~Debbah, ``Time reversal for {6G}
  wireless communications: {N}ovel experiments, opportunities, and
  challenges,'' \emph{IEEE Veh. Technol. Mag.}, vol.~17, no.~4, pp. 74--82,
  Dec. 2022.

\bibitem{montaldo2004telecommunication}
G.~Montaldo, G.~Lerosey, A.~Derode, A.~Tourin, J.~de~Rosny, and M.~Fink,
  ``Telecommunication in a disordered environment with iterative time
  reversal,'' \emph{Waves in Random Media}, vol.~14, no.~3, pp. 287--302, 2004.

\bibitem{mokh2022time}
A.~Mokh, J.~de~Rosny, G.~C. Alexandropoulos, R.~Khayatzadeh, M.~Kamoun,
  A.~Ourir, A.~Tourin, and M.~Fink, ``Time reversal for multiple access and
  mobility: Algorithmic design and experimental results,'' in \emph{Proc. IEEE
  Wireless Communications and Networking Conference (WCNC)}, Austin, USA, Apr.
  2022, pp. 1731--1736.

\bibitem{heath2018foundations}
R.~W. Heath~Jr and A.~Lozano, \emph{Foundations of MIMO communication}.\hskip
  1em plus 0.5em minus 0.4em\relax Cambridge University Press, 2018.

\bibitem{ligorria1999iterative}
J.~P. Ligorr{\'\i}a and C.~J. Ammon, ``Iterative deconvolution and
  receiver-function estimation,'' \emph{Bulletin of the seismological Society
  of America}, vol.~89, no.~5, pp. 1395--1400, 1999.

\end{thebibliography}


\end{document}